\documentclass[11pt]{article}

\usepackage{amssymb,amsmath,amsthm,color,url,epsfig}

\newcommand\x{{\bf x}}
\newcommand\y{{\bf y}}
\newcommand\z{{\bf z}}
\newcommand\bfa{{\bf a}}
\newcommand\bfb{{\bf b}}
\newcommand\bfc{{\bf c}}

\newcommand\bff{{\bf f}}

\newcommand\bfs{{\bf s}}
\newcommand\bft{{\bf t}}

\newcommand\bfv{{\bf v}}
\newcommand\bfw{{\bf w}}
\newcommand\zero{{\bf 0}}
\newcommand\cc{{\mathbb C}}
\newcommand\zz{{\mathbb Z}}

\newtheorem{theorem}{Theorem}[section]

\newtheorem{example}[theorem]{Example}
\newtheorem{algorithm}[theorem]{Algorithm}
\newtheorem{remark}[theorem]{Remark}
\newtheorem{definition}[theorem]{Definition}
\newtheorem{proposition}[theorem]{Proposition}

\begin{document}

\title{A Polyhedral Method to compute all Affine Solution Sets
       of Sparse Polynomial Systems\thanks{This material is based
upon work supported by the National Science Foundation under
Grant No.\ 1115777.}}

\author{Danko Adrovic and Jan Verschelde\\
Department of Mathematics, Statistics, and Computer Science \\
University of Illinois at Chicago \\
851 South Morgan (M/C 249) \\
Chicago, IL 60607-7045, USA \\
Emails: \texttt{adrovic@math.uic.edu} and \texttt{jan@math.uic.edu} \\
URLs: \texttt{\url{www.math.uic.edu/~adrovic}} and
\texttt{\url{www.math.uic.edu/~jan}}}

\date{14 October 2013}

\maketitle

\begin{abstract}
To compute solutions of sparse polynomial systems efficiently
we have to exploit the structure of their Newton polytopes.
While the application of polyhedral methods naturally excludes
solutions with zero components, an irreducible decomposition of
a variety is typically understood in affine space, 
including also those components with zero coordinates.
We present a polyhedral method to compute all affine solution sets
of a polynomial system.  The method enumerates all factors contributing 
to a generalized permanent.
Toric solution sets are recovered as a special case of this enumeration.
For sparse systems as adjacent 2-by-2 minors our methods scale much
better than the techniques from numerical algebraic geometry.

\end{abstract}

\section{Introduction}

Given is $\bff(\x) = \zero$, a polynomial system of $N$ polynomials
$\bff = (f_1,f_2,\ldots,f_N)$ in $n$ unknowns $\x = (x_1,x_2,\ldots,x_n)$.
We assume our polynomials are {\em sparse} and only few (relative to
the degrees) monomials appear with nonzero coefficient.
The structure of sparse polynomials in several variables is captured
by their Newton polytopes.
A {\em polyhedral method} exploits the structure of the Newton polytopes
to efficiently compute the solutions of the polynomial system.
For the problem of computing solution sets in the intersection of
some coordinate planes, the direct application of a polyhedral method
fails, because the Newton polytopes change drastically
when selected variables become zero.

If every polynomial in the system has the same Newton polytope,
then the volume of that Newton polytope bounds the number of isolated
solutions with nonzero coordinates, as proven in~\cite{Kus76}.
This theorem was generalized in~\cite{Ber75}
and its constructive proof was implemented in~\cite{VVC94}.
A more general algorithm was given in~\cite{HS95}.
The problem of counting the number of isolated solutions in affine space
was first addressed in~\cite{Roj94}, see also~\cite{Roj99}, and~\cite{RW96}.
Stable mixed volumes, introduced in~\cite{HS97},
give an upper bound on the number of isolated solutions in affine space.
Methods to compute stable mixed volumes efficiently were proposed 
in~\cite{EV99} and~\cite{GLW99}.

The complexity of counting the total number of
affine solutions of a system of $n$ binomials (two monomials with
nonzero coefficients) in~$n$ variables was shown as \#P-complete~\cite{CD07}.
In \cite{HJS12} combinatorial conditions for the existence of 
positive dimensional solution sets are given,
for use in a geometric resolution~\cite{GLS01}.

Finiteness results in celestial mechanics were proven with
polyhedral methods in~\cite{HM06}, \cite{JH11}.
Tropical algebraic geometry, see e.g.:~\cite{BJSST07},
and in particular the fundamental theorem by~\cite{JMM08}, \cite{Pay09},
provides inspiration for a polyhedral computation of
all positive dimensional solution sets.  
In \cite{Ver09}, \cite{AV11,AV12,AV13}, Puiseux series were proposed 
to develop solution sets of polynomial systems, starting at infinity.
Coordinate transformations, similar to the ones in~\cite{AV12},
were applied in a more general setting in~\cite{HL12}.
For parametric binomial systems, an algorithm
(using the Smith normal form) of polynomial complexity
for the solutions with nonzero values of the variables
was presented in~\cite{GW12}.

One of the earliest descriptions of software to compute a primary
decomposition of binomial ideals
were published in~\cite{BLR99} and~\cite{OP00}.
Recent algebraic algorithms are in~\cite{Kah10}.
In relation to the general binomial primary decomposition of~\cite{ES96},
our motivation stems from Puiseux series (over~$\cc$)
and the ideals we obtain are radical.

Our first contribution is to formulate the search for affine solution
sets as the enumeration of all factors that contribute to a generalized
permanent.  This enumeration extends directly to general sparse systems.
Our second contribution is a polyhedral method to
compute Puiseux series for all affine solution sets.
Thirdly, prototypes for the proposed algorithms are implemented
in PHCpack in~\cite{Ver99}.  We tested our methods and software
on the family of adjacent 2-by-2 minors,
a problem described in~\cite{DES98} and~\cite{HS00}.
For such sparse systems, our method scales much better than the
techniques of numerical algebraic geometry~\cite{BDHPPSSW12}.

\section{Monomial Parametrizations of Affine Solution Sets}

Toric ideals are introduced with monomial maps
in~\cite[Chapter~4]{Stu96}.
In this section we define the representations of solution sets of
binomial systems and give examples to illustrate the difference between
the toric and the affine case.

Monomials $x_1^{a_1} x_2^{a_2} \cdots x_n^{a_n}$
in $n$ variables $\x = (x_1,x_2,\ldots,x_n)$ are defined
by exponents $\bfa = (a_1,a_2,$ $\ldots$, $a_n) \in \zz^n$ and
abbreviated as~$\x^\bfa$.  Denote $\cc^* = \cc \setminus \{ 0 \}$.
A binomial system consists of equations 
$c_\bfa \x^\bfa - c_\bfb \x^\bfb = 0$,
with $c_\bfa, c_\bfb \in \cc^*$ and $\bfa, \bfb \in \zz^n$.
If we are interested in {\em toric solutions}, i.e.: $\x \in (\cc^*)^n$,
then we normalize the equation $c_\bfa \x^\bfa - c_\bfb \x^\bfb = 0$
into $\x^{\bfa-\bfb} = c_\bfb/c_\bfa$. 
After normalization, we write a binomial system as $\x^A = \bfc$, 
where the matrix $A$ collects the exponent vectors and
the coefficient vector $\bfc$ stores the coefficients.
The null space of~$A$ gives exponent vectors for a unimodular coordinate
transformation leading to a monomial parametrization of the solution set.

\begin{definition} \label{defmonpar}
{\rm A {\em monomial parametrization} of 
a $d$-dimensional solution set in~$\cc^n$ is
\begin{equation} \label{eqmonpar}
   x_k = c_k t_1^{v_{1,k}} t_2^{v_{2,k}} \cdots t_d^{v_{d,k}},
   \quad c_k \in \cc^*, v_{i,k} \in \zz,
   \mbox{ for } i=1,2,\ldots,d \mbox{ and } k=1,2,\ldots,n.
\end{equation}
}
\end{definition}

Substituting (\ref{eqmonpar}) into $\x^\bfa$ we can write
$\x^\bfa = c_1^{a_1} c_2^{a_2} \cdots c_n^{a_n}
             t_1^{\langle \bfa , \bfv_1 \rangle}
             t_2^{\langle \bfa , \bfv_2 \rangle} \cdots
             t_d^{\langle \bfa , \bfv_d \rangle}$
where $\langle \bfa , \bfv_i \rangle 
= a_1 v_{i,1} + a_2 v_{i,2} + \cdots + a_n v_{i,n}$.
Because
the $d$ vectors $\bfv_i$, $i=1,2,\ldots,d$, span a basis for the null
space of the vectors $\bfa - \bfb$ of the binomial equations
$c_\bfa \x^\bfa - c_\bfb \x^\bfb = 0$, we have
$d$ free parameters in $\bft = (t_1,t_2,\ldots,t_d)$.
Collecting the $d$ vectors $\bfv_i$ into the columns of an
$n$-by-$d$ matrix~$V$, and the coefficients $c_k$ of~(\ref{eqmonpar})
in the vector~$\bfc_V$, we abbreviate a  monomial parametrization 
in~(\ref{eqmonpar}) as $\x = \bfc_V \bft^V$.

\begin{example} \label{exunimonpar}
{\rm Because we want invertible coordinate transformations,
we may need fractional powers.
\begin{equation}
   \begin{array}{c}
      \bff(\x) = 
      \left\{
         \begin{array}{l}
            x_1^{80} - x_2^{21} x_3^2 = 0 \\ \vspace{-3mm} \\
            x_1^{54} - x_2^{15} x_4^2 = 0 \\
         \end{array}
      \right. \\ \vspace{-3mm} \\
      A =
      \left[
         \begin{array}{cccc}
           -80 & 21 & 2 & 0 \\
           -54 & 15 & 0 & 2 
         \end{array}
      \right] 
   \end{array}
   \quad
   M = 
   \left[
      \begin{array}{cccc}
         5 &  21/22 & 0 & 0 \\
        18 &  80/22 & 0 & 0 \\
        11 &    0   & 1 & 0 \\
         0 & -33/22 & 0 & 1 \\
      \end{array}
   \right]
   \quad
   \left\{
      \begin{array}{l}
         x_1 = y_1^5 y_2^{21/22} \\ \vspace{-3mm} \\
         x_2 = y_1^{18} y_2^{80/22} \\ \vspace{-3mm} \\
         x_3 = y_1^{11} y_3 \\ \vspace{-3mm} \\
         x_4 = y_2^{-33/22} y_4 \\
      \end{array}
   \right.
\end{equation}
The first two columns of~$M$ span the null space of~$A$.
The denominator~22 is obtained from the pivot of the Hermite
normal form of an integer vector that spans the null space of~$A$.
Dividing the columns of $B$ by these pivots gives an extended
matrix~$M$ with $\det(M) = 1$.
We denote the coordinate transformation defined by~$M$ as $\x = \y^M$.
Because $M$ contains the null space of~$A$,
$\bff(\x = \y^M) = \zero$ contains only $y_3$ and $y_4$ 
(after clearing the common powers of $y_1$ and $y_2$).
Solving $\bff(\y) = \zero$ gives the coefficients
$c_3 = \pm 1$ and $c_4 = \pm 1$ we put in place for $y_3$ and $y_4$:
$(x_1 = t_1^5 t_2^{21/22},
  x_2 = t_1^{18} t_2^{80/22},
  x_3 = (\pm 1) t_1^{11},
  x_4 = (\pm 1) t_2^{-33/22})$,
or equivalently:
$(x_1 = s_1^5 s_2^{21},
  x_2 = s_1^{18} s_2^{80},
  x_3 = (\pm 1) s_1^{11},
  x_4 = (\pm 1) s_2^{-33},
  t_1 = s_1,
  t_2 = s_2^{22})$.
By the auxiliary parameters~$s_1$ and~$s_2$,
the parametrization has integer exponents.  }
\end{example}

For a correct determination of the degree of the solution set,
we need unimodular monomial parametrizations.
Using the abbreviated notation $\x = \bfc_V \bft^V$,
we extend Definition~\ref{defmonpar}.

\begin{definition} {\rm A {\em unimodular monomial parametrization}
of a $d$-dimensional solution set in $\cc^n$ is
\begin{equation}
   (\x , \bft) = (\bfc_V \bfs^V, \bfs^W) \quad \mbox{ or } \quad
   (\x  = \bfc_V \bfs^V, \bft = \bfs^W),
\end{equation}
where $V \in \zz^{n \times d}$ and $W \in \zz^{d \times d}$.
The columns of~$V$ span the null space of the exponent vectors of 
the binomials and $W$ is a diagonal matrix containing the denominators 
of the columns of $V$ so that when $V W^{-1}$ 
is extended with unit vectors into the
square matrix~$M$, $\det(M) = 1$ and~$\x = \y^M$ is unimodular. }
\end{definition}

We assume that all our monomial parametrizations are unimodular
and omit~$\bfs$ when $W = I$.

An {\em affine solution set} is a component of a solution set
contained in a subspace spanned by one or more coordinate hyperplanes.
Some coordinates of an affine solution are zero, some are free,
and others are linked to a toric solution of a subset of the
original equations.
We illustrate the distinction between variables in the next example
and make this distinction precise in Definition~\ref{defaffinesolrep},
extending Definition~\ref{defmonpar} one last time.

\begin{example} {\rm 
An interesting class of examples are the adjacent minors
(see~\cite{DES98,HS00,Stu02}).
Consider all adjacent 2-by-2 minors of a general 2-by-4 matrix~$X$:
\begin{equation}
   X = 
   \left[
      \begin{array}{cccc}
         x_{11} & x_{12} & x_{13} & x_{14} \\
         x_{21} & x_{22} & x_{23} & x_{24} \\
      \end{array}
   \right]
   \quad \quad
   \bff(\x) = 
   \left\{
      \begin{array}{c}
         x_{11} x_{22} - x_{21} x_{12} = 0 \\
         x_{12} x_{23} - x_{22} x_{13} = 0 \\
         x_{13} x_{24} - x_{23} x_{14} = 0 \\
      \end{array}
   \right.
\end{equation}
which has a 5-dimensional toric solution
$(x_{11} = t_1 t_4 t_5,
  x_{12} = t_2,
  x_{13} = t_3,
  x_{14} = t_5,
  x_{22} = t_2 t_4^{-1} t_5^{-1}$,
 $x_{21} = t_1,
  x_{23} = t_3 t_4^{-1} t_5^{-1},
  x_{24} = t_4^{-1})$
of degree four and two affine solutions, each of degree two.
Giving the variables in the third column of~$X$ the value zero
reduces the original system to one equation.
The variables $x_{14}$ and $x_{24}$ no longer occur in the
remaining equations and are free.  
The other variables are interlinked.
For each variable in the solution we explicitly indicate its type:
$(x_{11} = t_1 t_2 t_3 {\rm (link)},
  x_{12} = t_3 {\rm (link)},
  x_{13} = 0 {\rm (zero)},
  x_{14} = t_4 {\rm (free)},
  x_{21} = t_2 {\rm (link)},
  x_{22} = t_1^{-1} {\rm (link)},
  x_{23} = 0  {\rm (zero)}, 
  x_{24} = t_5  {\rm (free)})$.
The other affine solution with $x_{12} = 0$ and $x_{22} = 0$
is obtained by symmetry.  }
\end{example}

\begin{definition} \label{defaffinesolrep}
{\rm An {\em affine monomial parametrization} of a solution set
of a binomial system is a tuple associating to the variables one
of the three types, {\em zero}, {\em free}, or {\em link}:
\begin{equation}
   x_k =
   \left\{
      \begin{array}{rcl}
         0 & & zero \\
         t_k & & free \\
         c_k \bft^\bfv & & link
      \end{array}
  \right.
\end{equation}
where $c_k \in \cc^*$,
$\bft = (t_1,t_2,\ldots,t_d)$, $d$ is the size of the
set of parameters that control the link variables, $\bfv \in \zz^d$, and
$\bft^\bfv = t_1^{v_1} t_2^{v_2} \cdots t_d^{v_d}$.
In particular, the distinction between a free and a link variable
is that the parameter $t_k$ does not occur anywhere else in the
monomial parametrization of the affine solution. }
\end{definition}

\begin{proposition}  \label{propdegaffset}
The degree of a solution component of dimension $D$ of a binomial system
given by an affine unimodular monomial parametrization equals the volume
of the polytope spanned by the origin and the exponent vectors of all 
parameters.  This polytope is described as follows.
Let $D = d+e$, where $d$ is the number of parameters
that control the link variables and $e$ is the number of free variables.
Relabel the free variables so they come before the link variables.
Then for every free variable $k$ we have the $k$th standard basis
vector ${\bf e}_k \in \zz^D$ and insert $e$ zeros to each $\bfv \in \zz^d$.
\end{proposition}

\noindent {\em Proof.}  To compute the degree of a $D$-dimensional 
solution set of $\bff(\x) = \zero$, we add $D$ generic hyperplanes 
$L(\x) = \zero$ and count the number of isolated solutions 
of~$\bff(\x) = \zero$ that satisfy $L(\x) = \zero$.  
By the monomial parametrization, we can eliminate the original $\x$ 
variables, omit the original equations $\bff(\x) = \zero$, and consider 
only the system $L(\bft) = \zero$.  
As the coefficients of the hyperplanes in~$L$ are generic,
all equations have the same monomials and exponents:
$L(\bft) = \zero$ has $D$ equations in $D$ unknowns.
The theorem of~\cite{Kus76} applies and the number of isolated solutions 
of~$L(\bft) = \zero$ equals the volume of the Newton polytope spanned by
the exponents of the polynomials in~$L(\bft) = \zero$.~\qed

\begin{remark} {\rm For ease of notation, we assumed $W = I$ in
Proposition~\ref{propdegaffset}.
If $W \not= I$, then the volume of the polytope
must be divided by~$\det(W)$ to obtain the correct degree.
For example, the solutions of Example~\ref{exunimonpar}
have degree~$54 = 1188/22$.
Proposition~\ref{propdegaffset} is similar to~\cite[Theorem~4.16]{Stu96}. }
\end{remark}

\section{A Generalized Permanent}

To enumerate all choices of variables to be set to zero,
we use the matrix of exponents of the monomials to define
a bipartite graph between monomials and variables.
The incidence matrix of this bipartite graph is defined below.

\begin{definition}  {\rm Let $\bff(\x) = \zero$ be a system.
We collect all monomials $\x^\bfa$ that occur in $\bff$ along
the rows of the matrix, yielding the {\em incidence matrix}
\begin{equation}
   M_\bff[\x^\bfa,x_k] = 
   \left\{
      \begin{array}{lcl}
         1 & {\rm if} & a_k > 0 \\
         0 & {\rm if} & a_k = 0. \\
      \end{array}
   \right.
\end{equation}
Variables which occur anywhere with a negative exponent
are dropped. }
\end{definition}

\begin{example} {\rm For all adjacent minors of a 2-by-3 matrix,
the matrix linking monomials to variables is
\begin{equation}
   M_\bff = 
   \left[
\begin{array}{c|ccccccc}
                & x_{11} & x_{12} & x_{13} & x_{21} & x_{22} & x_{23} \\ \hline
  x_{11} x_{22} &   1    &   0    &   0    &   0    &   1    &   0    \\
  x_{21} x_{12} &   0    &   1    &   0    &   1    &   0    &   0    \\
  x_{12} x_{23} &   0    &   1    &   0    &   0    &   0    &   1    \\
  x_{22} x_{13} &   0    &   0    &   1    &   0    &   1    &   0    \\
\end{array}
   \right],
   \quad \mbox{for} \quad
   \bff = \left\{
      \begin{array}{l}
         x_{11} x_{22} - x_{21} x_{12} = 0 \\
         x_{12} x_{23} - x_{22} x_{13} = 0. \\
      \end{array}
   \right.
\end{equation}
For this example, the rows of $M_\bff$ equal the exponents of the monomials.
We select $x_{12}$ and $x_{22}$ as variables to be set to zero,
as overlapping columns $x_{12}$ with $x_{22}$ gives all ones. }
\end{example}

\begin{proposition}
Let $S$ be a subset of variables such that
for all $\x^\bfa$ occurring in $\bff(\x) = \zero$: $M[\x^\bfa,x_k] = 1$, 
for $x_k \in S$, then setting all $x_k \in S$ to zero
makes all polynomials of $\bff$ vanish.
\end{proposition}
 
\noindent {\em Proof.} 
$M[\x^\bfa,x_k] = 1$ means: $x_k = 0 \Rightarrow \x^\bfa = 0$.
If the selection of the variables in~$S$ is such that all monomials
in the system have at least one variable appearing with positive power,
then setting all variables in~$S$ to zero makes all monomials in the
system vanish.~\qed

Enumerating all subsets of variables so that $\bff$ vanishes 
when all variables in a subset are set to zero is 
similar to a row expansion algorithm on~$M_\bff$ for a permanent,
sketched in Algorithm~\ref{algenumerate}.

\begin{algorithm}[recursive subset enumeration via row expansion of permanent]
\label{algenumerate} {\rm

\begin{tabbing} \\
\hspace{1cm} \= Input: \= $M_\bff$ is the incidence matrix 
                          of $\bff(\x) = \zero$; \\
             \>        \> index of the current row in $M_\bff$;
                          and $S$ is the current selection of variables. \\
             \> Output: all $S$ that make the entire $\bff$ vanish. \\
 \> if \= $M[\x^\bfa,x_k] = 1$ for some $x_k \in S$ \\
 \>    \> then \= print $S$ if $\x^\bfa$ is at the last row of $M_\bff$ 
                            or else go to the next row \\
 \>    \> else \> for \= all $k$: $M[\x^\bfa,x_k] = 1$ do \\
 \>    \>      \>     \> $S := S \cup \{ x_k \}$ \\
 \>    \>      \>     \> print $S$ if $\x^\bfa$ is 
                         at the last row of $M_\bff$ 
                         or else go to the next row \\
 \>    \>      \>     \> $S := S \setminus \{ x_k \}$
\end{tabbing}
}
\end{algorithm}

To limit the enumeration, every variable set to zero cuts the dimension 
of the solution set by one.
If we have a threshold on the dimension of the solution set,
then the enumeration stops if the number of selected variables
exceeds the threshold on the codimension.
In the context of algebraic sets, a greedy enumeration should first
search for the highest dimensional components and taking into account
the frequencies of the variables occurring in each monomial,
select the most frequently occurring variables first.

The above algorithm returns subsets of variables that make the
entire binomial system vanish.  For partial cancellation,
note that skipping certain binomials 
means skipping pairs of rows in $M_\bff$.
The odd (respectively even) rows of
$M_\bff$ store the first (respectively second) monomial.
Then the extra branch test in the enumeration proceeds as follows.
If the current row in $M_\bff$ is odd and
if none of the selected variables occur in the current $\x^\bfa$
and in $\x^\bfb$ on the following row,
then skip the row in one branch of the enumeration.
Skipping one binomial equation implies that variables
occurring with positive power in $\x^\bfa$ and $\x^\bfb$
should not be selected in the future.
The skipped binomial equations define a toric solution for some variables
in an affine monomial parametrization.

\section{Membership Tests}

Regardless of efficient greedy enumeration strategies,
we still need a criterion to decide whether no member
of a collection of affine monomial parametrizations 
is contained in another parametrization.
We introduce the problem with an example.

\begin{example} \label{exhierarchies}
{\rm To illustrate the hierarchies of variables
and monomials when skipping equations we consider the system
taken from~\cite[Example~2.2]{HS00}:
\begin{equation} \label{eqexample2_2}
  \bff(\x) = 
  \left\{
     \begin{array}{l}
        x_1 x_3^2 - x_2 x_6^2 = 0 \\
        x_4 x_6^3 - x_1^3 x_5 = 0 \\
        x_1 x_2 x_5 - x_4 x_6^2 = 0.
     \end{array}
  \right.
\end{equation}
The system has two 3-dimensional toric solution sets:
$(x_1 = t_1^2 t_2^2 t_3,
  x_2 = t_1^4 t_2^4 t_3,
  x_3 = \pm t_1 t_2  t_3,
  x_4 = t_1^6,
  x_5 = t_2^{-6},
  x_6 = t_3)$,
one 4-dimensional affine solution set:
$(x_1 = 0, x_2 = t_1, x_3 = t_2, x_4 = t_3, x_5 = t_4, x_6 = 0)$,
and three 3-dimensional affine solution sets:
$(x_1 = 0, x_2 = 0, x_3 = t_1, x_4 = 0, x_5 = t_2, x_6 = t_3)$,
$(x_1 = t_1, x_2 = t_2, x_3 = t_3, x_4 = 0, x_5 = 0, x_6 = 0)$,
and
$(x_1 = t_1 t_2^2 t_3^2, x_2 = t_1, x_3 = t_2^{-1}, x_4 = 0, x_5 = 0,
  x_6 = t_3)$. 
In the enumeration, after $x_1 = 0$ and $x_6 = 0$ have been found
to completely set the system to zero, any additional sets of variables
that include $x_1$ and $x_6$ should no longer be considered.
Variables $x_1$ and $x_6$ occur most often in the system and if we order
the variables along their frequency of occurrence, then $x_1 = 0$
and $x_6 = 0$ will be considered first, before all other pairs, and
subsets containing $x_1$ and $x_6$.  }
\end{example}

Two equivalent (as defined below) monomial parametrizations 
describe the same solution set.

\begin{definition} {\rm Consider two monomial parametrizations
$\bfc_V \bft^V$ and $\bfc_W \bft^W$.  
If $\bfc_V = \bfc_W$ and the matrices $V$ and $W$
span the same linear space, then we say that the monomial parametrizations
$\bfc_V \bft^V$ and $\bfc_W \bft^W$ are {\em equivalent}.
Two affine monomial parametrizations are {\em equivalent} if the same
variables are zero, the same variables are free, and moreover,
their link variables have equivalent monomial parametrizations. }
\end{definition}

We have to be able to decide whether an affine monomial representation
belongs to another (affine) monomial parametrization.
We introduce this problem in the following example.

\begin{example} [Example~\ref{exhierarchies} continued]
{\rm Consider~(\ref{eqexample2_2}).  The enumeration generates
$C_1 = (x_1 = t_1, x_2 = 0, x_3 = 0, x_4 = 0, x_5 = 0, x_6 = t_2)$.
But as it turns out, this component is a subset of
$C_2 = (x_1 = t_1 t_2^2 t_3^2, x_2 = t_1, x_3 = t_2^{-1}, x_4 = 0, x_5 = 0,
  x_6 = t_3)$.  This is not at all obvious from the given
parametrization of~$C_2$ because $x_3$ cannot become zero in~$C_2$
because of the negative power of~$t_2$.  With some manipulations,
we can find an equivalent parametrization for~$C_2$:
$(x_1 = t_1, x_2 = t_1 t_2^2, x_3 = t_2 t_3, x_4 = 0, x_5 = 0, x_6 = t_3)$
and then $t_2 = 0$ leads to $C_1$.  
A better way to consider whether $C_1 \subseteq C_2$ is
to observe that $C_1$ is a monomial ideal, that is: the ideal $I(C_1)$ 
defined by all polynomials that vanish at $C_1$ is generated by
$\langle x_2, x_3, x_4, x_5 \rangle$.  
We have $I(C_2) = \langle x_4, x_5, x_1 x_3^2 - x_2 x_6^2 \rangle$.
Comparing $I(C_2)$ with $I(C_1)$, we observe that
$x_1 x_3^2 - x_2 x_6^2 = (x_1 x_3) x_3 + (-x_6^2) x_2 \in I(C_1)$
and thus $I(C_2) \subseteq I(C_1)$, which implies $C_1 \subseteq C_2$.
Verifying whether a polynomial belongs to a monomial ideal seems
easier than finding an equivalent parametrization with positive
powers at the right places.  }
\end{example}

Given an affine monomial parametrization $C$ of a solution set $V(C)$
of a binomial system, the ideal $I(C)$ of all polynomials that vanish
at~$V(C)$ consists of monomials (defined by those variables that are
set to zero) and binomial relations (defined by the power products in
the parametrizations).   
The monomial parametrizations of the affine solution sets remove
the multiplicities, e.g.: $(x - y)^2$ turns into $x-y$, and we therefore
have that $V(I(C)) = C$, for any affine component~$C$.

\begin{algorithm}[defining equations via circuit enumeration]
\label{algequcirc} {\rm

\begin{tabbing} \\
\hspace{1cm} \= Input: $C$ an affine monomial map of solutions $V(C)$. \\
             \> Output: $E(C)$ equations that define $V(C)$.
\end{tabbing}
}
\end{algorithm}
The implementation of Algorithm~\ref{algequcirc} considers
all smallest affine dependencies of the exponents of the monomials.
A smallest affine dependency between points 
is a circuit~\cite{BLSWZ99}.

In the proposition below we formalize the ideal inclusion property
according to our notations.

\begin{proposition} \label{propinclusion}
Let $C_1$ and $C_2$ be two affine monomial parametrizations
for solutions sets of~$\bff(\x) = \zero$.
Then $C_1 \subseteq C_2 \Leftrightarrow I(C_1) \supseteq I(C_2)$.
\end{proposition}

Although the enumeration of all equations that define~$V(C)$
has once again a combinatorial complexity, often only one
particular equation solves the inclusion problem, illustrated next.

\begin{example} [Example~\ref{exhierarchies} continued]
{\rm Could a toric component include
the set defined by $x_4 = 0$ and $x_5 = 0$?
To answer this question, it suffices to consider coordinates of
the toric component  that do not involve~$x_4$ and $x_5$, for example:
$x_1 = t_1^2 t_2^2 t_3, x_2 = t_1^4 t_2^4 t_3, x_6 = t_3$.
The monomials in the parameters define the exponent matrix~$A$
and its null space defines a vanishing binomial:
\begin{equation}
   A = \left[
      \begin{array}{ccc}
         2 & 2 & 1 \\
         4 & 4 & 1 \\
         0 & 0 & 1 \\
      \end{array}
   \right],
   \quad A \left[
              \begin{array}{r}
                 2 \\ -1 \\ -1
              \end{array}
           \right]
   = \zero,
   \quad x_1^2 - x_2 x_6 = 0, 
   \quad x_1^2 - x_2 x_6 \not\in \langle x_4, x_5 \rangle.
\end{equation}
Therefore,
the set with $x_4 = 0$ and $x_5 = 0$ does not belong to the set
defined by $x_1^2 - x_2 x_6 = 0$.  }
\end{example}

Summarizing the properties of $M_\bff$ and 
Proposition~\ref{propinclusion}, we state that
all irreducible components of the solution set of
a binomial system $\bff(\x) = \zero$ are factors contributing to 
the generalized permanent of the incidence matrix~$M_\bff$.
Moreover, the affine parametrizations of the components give enough
equations to determine that every component reported by the enumeration
does not belong to any other component.

\section{Enumerating All Candidate Affine Solution Sets}

To find all candidate affine solution sets,
we sketch an extension of Algorithm~\ref{algenumerate}.

\begin{example}[Our running example] \label{exrunning}
{\rm Consider~\cite[example~8]{HJS12}:
\begin{equation} \label{eqexrunning}
  \left\{
  \begin{array}{r}
     x_1 x_4 + x_1^2 x_4^2 + x_1 x_2 x_3 + x_2 x_3 = 0 \\
    \vspace{-3mm} \\
     x_1 x_2 + x_1 x_2^2 + x_1 x_3 x_4 + x_3 x_4 + x_3 x_4^2 = 0 \\
    \vspace{-3mm} \\
     x_1 x_2 x_4 + x_1 x_3 x_4 + x_2 x_3 + x_2 x_3 x_4 = 0 \\
    \vspace{-3mm} \\
     x_1 + x_1^2 + x_1 x_2 + x_3^2 + x_3 x_4 = 0
  \end{array}
  \right.
\end{equation}
where we have taken all coefficients to be equal to one.
Notice that $x_1$ and $x_3$ appear in every monomial, 
so setting $x_1$ and $x_3$ to zero yields a 2-dimensional solution set.}
\end{example}

If $x_1$ is set to zero, then also $x_1^2$ becomes zero,
so in the incidence matrix we consider only those monomials
which are not divided by any other monomial, as formalized
in the next definition.

\begin{definition} {\rm 
The supports of $\bff = (f_1,f_2,\ldots,f_N)$ 
are $(A_1,A_2,\ldots,A_N)$:
$\displaystyle f_i(\x) = \sum_{\bfa \in A_i} c_\bfa \x^\bfa$.
The {\em incidence matrix} for $f_i$ is $M_{f_i}$:
for all $\bfa \in A_i$ for which there is no 
$\bfb \in A_i \setminus \{ \bfa \}$ such that $\x^\bfb$ divides $\x^\bfa$:
\begin{equation}
   M_{f_i}[\x^\bfa,x_k] = \left\{
     \begin{array}{lcl}
        1 & {\rm if} & a_k > 0 \\
        0 & {\rm if} & a_k = 0 \\
     \end{array}
   \right.
   \quad \mbox{and} \quad
   M_\bff = \left[
     \begin{array}{c|c|c|c}
        M_{f_1} & M_{f_2} & \cdots & M_{f_N}
     \end{array}
   \right]^T.
\end{equation}
}
\end{definition}

\begin{example} [Example~\ref{exrunning} continued]  {\rm
The incidence matrix for~(\ref{eqexrunning}) is
\begin{equation}
  M_\bff =
  \left[
     \begin{array}{ccc|ccc|cccc|cccc}
        1 & 1 & 0   &   1 & 1 & 0   &   1 & 1 & 0 & 0   &    1 & 1 & 0 & 0 \\
        0 & 1 & 1   &   1 & 0 & 0   &   1 & 0 & 1 & 1   &    0 & 1 & 0 & 0 \\
        0 & 1 & 1   &   0 & 1 & 1   &   0 & 1 & 1 & 1   &    0 & 0 & 1 & 1 \\
        1 & 0 & 0   &   0 & 1 & 1   &   1 & 1 & 0 & 1   &    0 & 0 & 0 & 1 \\
     \end{array}
  \right]^T.
\end{equation}
Observe that the transposed matrix is displayed.
The rows of $M_\bff^T$ are indexed by the variables.
}
\end{example}

Running through the columns of~$M_\bff$ seems equivalent
to enumerating all subsets of~$\{ x_1, x_2$, $\ldots$, $x_n \}$.
Organizing the search along the rows of~$M_\bff$ allows
for a greedy version, see Figure~\ref{figgreedysearch}.
For example, we could first set those
variables to zero which appear most frequently in the monomials.
Running Algorithm~\ref{algenumerate} through all equations, 
we obtain solutions that make all equations of~$\bff$ vanish.

\begin{figure}[ht]
\begin{center}
\begin{picture}(230,60)(0,0)
\put(0,20){
\begin{picture}(200,100)(0,0)
\put(13,23){\textcolor{red}{\circle{12}}}
\put(12,29){\textcolor{red}{\vector(1,0){205}}}
\put(12,17){\textcolor{red}{\vector(1,0){205}}}
\put(29,-3){\textcolor{red}{\circle{12}}}
\put(29,3){\textcolor{red}{\vector(1,0){188}}}
\put(29,-9){\textcolor{red}{\vector(1,0){188}}}
\put(-30,0){$\displaystyle
  M_\bff = \left[
     \begin{array}{ccc|ccc|cccc|cccc}
        1 & 1 & 0   &   1 & 1 & 0   &   1 & 1 & 0 & 0   &    1 & 1 & 0 & 0 \\
        0 & 1 & 1   &   1 & 0 & 0   &   1 & 0 & 1 & 1   &    0 & 1 & 0 & 0 \\
        0 & 1 & 1   &   0 & 1 & 1   &   0 & 1 & 1 & 1   &    0 & 0 & 1 & 1 \\
        1 & 0 & 0   &   0 & 1 & 1   &   1 & 1 & 0 & 1   &    0 & 0 & 0 & 1 \\
     \end{array}
  \right]^T$}
\end{picture}
}
\end{picture}
\caption{Searching greedily, we first select $x_1 = 0$.  
Then we look for the first monomial that does not contain $x_1$ 
and we choose $x_3$ over $x_2$ because $x_3$ appears in more monomials.}
\label{figgreedysearch}
\end{center}
\end{figure}

\section{A Polyhedral Method}

Skipping a binomial equation, e.g.:
$x_{11} x_{22} - x_{21} x_{12} \not= 0$
implies $x_{11} \not= 0$, $x_{22} \not= 0$,
$x_{21} \not= 0$, and $x_{12} \not= 0$.
For general polynomial equations, it suffices that 
{\em at least two} monomials remain.
A purely combinatorial criterion is to consider all possible binomials
to determine which variables should be nonzero.

In a polyhedral method we examine inner normals to determine
initial form systems.  For every skipped polynomial~$p$,
we consider all edges of its Newton polytope.
For each edge $e$ with inner normal cone $V$,
let ${\rm in}_V(p)$ be the initial form of~$p$:
${\rm in}_V(p)$ contains those terms $c_\bfa \x^\bfa$ of $p$ 
for which $\langle \bfa , \bfv \rangle$ is minimal for
all $\bfv$ in the interior of the cone~$V$.
Instead of the pure combinatorial criterion from above, we now require
that all variables occurring in ${\rm in}_V(p)$ should remain nonzero.
The intersection of the inner normal cones of equations
that are skipped determine the pretropism(s).
Our polyhedral method to enumerate all candidate affine
solution sets has input/output specification 
in Algorithm~\ref{alginputoutput}.

\begin{algorithm}[input/output specification of polyhedral method]
{\rm
\label{alginputoutput}
\begin{tabbing}
\\
\hspace{5mm} \= Input: \= $M_\bff$, 
             the incidence matrix of $\bff(\x) = \zero$.
             $E = (E_1, E_2, \ldots, E_N)$, \\
\>        \> $E_i$ is the set of all edges
             of the Newton polytope of $f_i$, $i=1,2,\ldots,N$. \\ 
\> Output: $S = \{ \ (s,e) \ | $ \=
                $n\mbox{-tuple}~s:$ 
                   $s_i = 0 \mbox{ if } x_i = 0$, 
            $s_i = +1 \mbox{ if } x_i \in \cc$, 
            $s_i = -1 \mbox{ if } x_i \not= 0$; and \\
\>            \>  $N\mbox{-tuple}~e:
                \ e_i = \emptyset \mbox{ or } e_i \in E_i,
                \mbox{ for } i=1,2,\ldots, N \ \}$.
\end{tabbing}
}
\end{algorithm}

Skipping all equations and making all edge-edge combinations
yields the refinement of normal cones for the tropical prevariety.
Normal cone intersections prune superfluous combinations.
Consider the processing of a tuple~$(s,e)$.
If the dimension of the normal cone defined by $e_i \not= \emptyset$
equals $D$, 
then we have $D$ parameters $t_1, t_2, \ldots, t_D$.
We have $D = \#\{ \ s_i = -1 \ | \ (s,e) \in S \ \}$.
For all $s_i = 1$, we have extra free variables $x_i = t_{D+j}$,
for $j = 1,2,\ldots,\#\{ \ s_i = +1 \ | \ (s,e) \in S \ \}$.

The specification of Algorithm~\ref{alginputoutput} fits the
description of the normal cone intersection algorithms to compute the tropical 
prevariety as defined by~\cite{BJSST07} and done by the software Gfan 
of~\cite{Jen08}.
Solutions to the initial form systems give the leading
powers of Puiseux series expansions.  Before we formalize the format
of these expansions, we continue our running example.

\begin{example}[Example~\ref{exrunning} continued] {\rm
There are five cases that lead to affine solution sets:

\vspace{-1mm}

\begin{enumerate}
\item Setting $x_{1}=0$ and $x_{2}=0$ leaves only
      $x_{3}x_{4}^{2} + x_{3}x_{4} = 0$ and $x_{3}^{2} + x_{3}x_{4} = 0$.
      The solutions are the line $(x_1 = 0, x_2 = 0, x_{3} = 0, x_{4} = t_1)$
      and $(0,0,1,-1)$.

\vspace{-1mm}

\item Setting $x_{2}=0$ and $x_{3}=0$ leaves only
      $x_{1}^{2}x_{4}^{2} + x_{1}x_{4} = 0$
      and $x_{1}^{2} + x_{1} = 0$.
      The solutions are the line
      $(x_{1} = 0, x_2 = 0, x_3 = 0, x_{4} = t)$,
      $(-1,0,0,0)$, and $(-1,0,0,1)$.

\vspace{-1mm}

\item Setting $x_{2}=0$ and $x_{4}=0$ leaves
      only $x_{1}^{2} + x_{3}^{2} + x_{1} = 0$.
      A Puiseux expansion for the solution starts as
      $(x_{1} = t^2(-1 + O(t^2)), x_2 = 0, 
        x_{3} = t(-1 + O(t^2)),x_4 = 0)$.

\vspace{-1mm}

\item Setting $x_{3}=0$ and $x_{4}=0$ leaves only
      $x_{1}x_{2}^{2} + x_{1}x_{2} = 0$
      and $x_{1}^{2} + x_{1}x_{2} + x_{1}$.
      The solutions are the line
      $(x_{1} = 0, x_{2} = t,x_3 = 0, x_4 = 0)$ and $(-1,0,0,0)$.

\vspace{-1mm}

\item Setting $x_{2}=0$, $x_{3}=0$ and $x_{4}=0$ leaves
      $x_{1}^2 + x_{1} = 0$, with solutions $(-1,0,0,0)$ and $(0,0,0,0)$.
\end{enumerate}

\vspace{-1mm} 

Stable mixed volumes 
will also lead to
all isolated solutions in affine space.  }
\end{example}

\begin{proposition} \label{propcommute}
Assume $\bff(\x) = \zero$ has an affine solution
set with $\ell$ link variables, $m$ free variables, 
and the remaining $n - \ell - m$ variables are set to zero.
Ordering variables so the link variables appear first, 
followed by the free and then the zero variables, we partition $\x$ as
\begin{equation}
   \x = (x_1,x_2,\ldots,x_\ell,
         x_{\ell+1}, x_{\ell+2}, \ldots, x_{\ell+m},
         x_{\ell+m+1}, x_{\ell+m+2}, \ldots, x_n).
\end{equation}
Ordering the parameters so the first $D$ parameters 
$t_1,t_2,\ldots,t_D$ occur in the link variables:
\begin{equation}
   \x = \left\{
     \begin{array}{ll}
        {\displaystyle x_k = c_k \prod_{j=1}^D t_j^{v_{k,j}}}(1+O(\bft))
                             & k=1,2,\ldots,\ell \\
        x_{\ell+k} = t_{D+k} & k=1,2,\ldots,m \\
        x_{\ell+m+k} = 0 & k = 1,2,\ldots,n-\ell-m
     \end{array}
   \right.
\end{equation}
with coefficients $c_k \in \cc^*$, $k=1,2,\ldots,\ell$ and
where the vectors $\bfv_1, \bfv_2, \ldots, \bfv_D \in \zz^k$ span
a $D$-dimensional cone~$V$.  
Take any $\bfv \in V$
and denote $\bfw = (\bfv,\zero,{\boldsymbol{\infty}})$,
where $\zero$ is a vector of $m$ zeros
and $\boldsymbol{\infty}$ a vector of $n-\ell-m$ infinite numbers.
Then, for $\z = (x_1,x_2,\ldots,x_\ell, t_{D+1},t_{D+2}, \ldots, t_{D+m}$,
$0, 0 ,\ldots, 0 )$:
\begin{equation} \label{eqcommute}
   {\rm in}_\bfv ( \bff(\z) ) = ( {\rm in}_\bfw \bff ) (\z),
   \mbox{ for all values } t_{D+k} \in \cc, k=1,2,\ldots,m.
\end{equation}
\end{proposition}

\noindent {\em Proof.}  To prove~(\ref{eqcommute}) we consider 
two cases.  When the last $n - \ell - m$ variables are zero,
either an equation vanishes entirely or some monomials remain.
For monomials $\x^\bfa$ in which no variable appears
with index larger than $\ell + m$, the inner product 
\begin{equation} \label{eqcommuteproof}
  \langle \bfa, \bfw \rangle 
   = a_1 v_1 + a_2 v_2 + \cdots + a_\ell v_\ell
   = \langle (a_1,a_2,\ldots,a_\ell), \bfv \rangle < \infty.
\end{equation}
For monomials in which there is at least one variable with index
larger than $\ell + m$, we have $\langle \bfa , \bfw \rangle = \infty$.

In the case where $f_i$ vanishes entirely when the last $n - \ell - m$ 
variables are zero, every monomial has at least one variable with 
index larger than $\ell + m$.
In that case ${\rm in}_\bfw (f_i) = f_i$ and $f_i(\z) = 0$.
As ${\rm in}_\bfv ( 0 ) = 0$, we have that (\ref{eqcommute}) holds for
all values $t_{D+k}$, $k=1,2,\ldots,m$.
In the other case, there are monomials $\x^\bfa$ 
in which no variables appear with index larger than $\ell + m$
and for those $\x^\bfa$: $\langle \bfa , \bfw \rangle < \infty$.
Vanishing monomials have a variable with index larger than
$\ell + m$ and $\langle \bfa , \bfw \rangle = \infty$.
By~(\ref{eqcommuteproof}), for monomials that have no variables
with index larger than~$\ell + m$, $\langle \bfa , \bfw \rangle$
equals the inner product with~$\bfv$.  
Thus~(\ref{eqcommute}) holds.~\qed

Proposition~\ref{propcommute} allows to make the connection with
stable mixed volumes.
In particular, the inner normals of the stable mixed cells contain
the origin that is lifted sufficiently high, leading to some
components in the inner normal of much higher magnitude than the others.
As in the case of those inner normals, we can extend the tropisms~$\bfv$
of the specialized system to tropisms~$\bfw$ of the original system,
where the values that correspond to the variables that are set to zero
are sufficiently high.

We end this paper with the observation that although most initial form
systems are not binomial, all Puiseux series have a leading term which 
satisfies a binomial system.  The combinatorial algorithms for the 
defining equations of monomial maps help solving the initial form systems.

\section{Computational Experiments}

Since version 2.3.68 of PHCpack~\cite{Ver99}, the black box solver 
(called as {\tt phc -b}) computes toric solutions of binomial systems.
This code is also available 
via the Python interface {\tt phcpy}~\cite{Ver13}.

The polynomial equations of adjacent minors
are defined in~\cite[page~631]{HS00}:
$x_{i,j} x_{i+1,j+1} - x_{i+1,j} x_{i,j+1} = 0$,
$i=1,2,\ldots,m-1$, $j = 1,2,\ldots,n-1$.
For $m=2$, the solution set is pure dimensional,
of dimension $2 n - (n-1) = n+1$, the number of irreducible
components of $X$ equals the $n$th Fibonacci 
number~\cite[Theorem~5.9]{Stu02}, 
and the degree of the entire solution set is $2^n$.
Knowing that the solution set is pure dimensional, our enumeration can
be restricted so only sets of the right dimension are returned:
for every variable we set to zero, one equation has to vanish as well.
With this assumption, our enumeration produces exactly the right number
of components. 

Table~\ref{tabadmbarplot} shows the comparison between the method
proposed in this paper and a witness set construction.  
This construction requires the computation of as many generic points
as the degree of the solution set, which is $2^{n-1}$ in this case.
For $n-1$ quadrics, a total degree homotopy is optimal in the
sense that no paths diverge.  Experimental results show that
the witness set construction has a limited range.
In addition, our new method returns the irreducible decomposition.

\begin{table}[hbt]
{\small
\begin{tabular}{r|r|r|r|r}
 $n$ & $2^{n-1}$ & \#maps & search & witness \\ \hline
  3  &        4  &      2 &   0.00 &   0.03 \\
  4  &        8  &      3 &   0.00 &   0.16 \\
  5  &       16  &      5 &   0.00 &   0.68 \\
  6  &       32  &      8 &   0.00 &   2.07 \\
  7  &       64  &     13 &   0.01 &   7.68 \\ 
  8  &      128  &     21 &   0.01 &  28.10 \\
  9  &      256  &     34 &   0.02 &  71.80 \\
 10  &      512  &     55 &   0.05 & 206.01 \\
 11  &     1024  &     89 &   0.10 & 525.46 \\
 12  &     2048  &    144 &   0.24 & ---~~~ \\
 13  &     4096  &    233 &   0.57 & ---~~~ \\
 14  &     8192  &    377 &   1.39 & ---~~~ \\
 15  &    16384  &    610 &   3.33 & ---~~~ \\
 16  &    32768  &    987 &   8.57 & ---~~~ \\
 17  &    65536  &   1597 &  21.36 & ---~~~ \\
 18  &   131072  &   2584 &  55.95 & ---~~~ \\
 19  &   262144  &   4181 & 140.84 & ---~~~ \\
 20  &   524288  &   6765 & 372.62 & ---~~~ \\
 21  &  1048576  &  10946 & 994.11 & ---~~~ \\
\end{tabular}
}
\caption{The construction of a witness set for all adjacent minors
of a general 2-by-$n$ matrix requires the tracking of $2^{n-1}$ paths
which is much more expensive than the combinatorial search.
For $n$ ranging from 3 to 21 we list times in seconds for the 
combinatorial search in column~3 and for the witness construction
in the last column, capping the time at 1000 seconds. }
\label{tabadmbarplot}
\end{table}

\begin{picture}(400,0)(0,0)
\put(190,100){\epsfig{figure=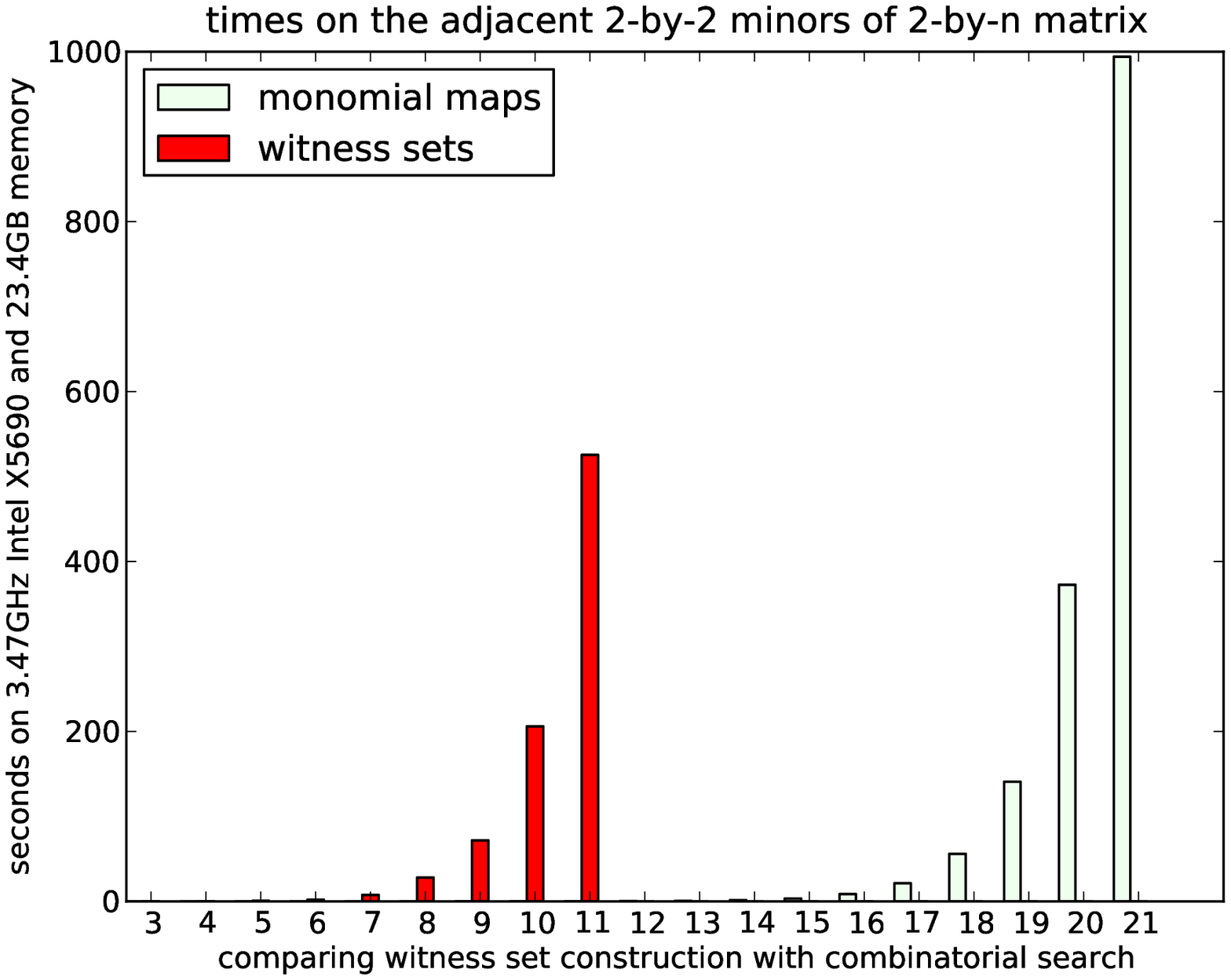,width=9cm}}
\end{picture}

The system of adjacent minors is also one of the benchmarks
in~\cite{BDHPPSSW12}, but neither Bertini~\cite{BHSW06} nor 
Singular~\cite{DGPS11} are able to get within the same range of PHCpack.
This is not a surprising conclusion since polyhedral methods
scale very well for binomial systems.

In~\cite[\S 5.3]{Stu02}, the adjacent minors introduce readers to the
joys of primary decomposition and the 4-by-4 case is explicitly
described in~\cite[Lemma~5.10]{Stu02} and~\cite[Theorem~5.11]{Stu02}.
Running {\tt phc~-b} we see 15 solution maps appear (in agreement
with the 15 primes of~\cite[Lemma~5.10]{Stu02}).  Of the 15, 12 maps
have dimension~9 and their degrees add up to~32.  There are two
linear solution sets of dimension~8 and one 7-dimensional solution
set of degree~20.  

All adjacent 2-by-2 minors of a general 5-by-5
matrix have 100 irreducible components.  There are two linear maps
of dimension 15, twelve 14-dimensional linear maps, 22 maps of dimension~13
with degrees adding up to~110, the sum of the degrees of the 63
12-dimensional maps equals 582, and finally, there is one 9-dimensional
solution set of degree 70.  The sum of the degrees of all 100 components
equals~776.


\end{document}